\documentclass{article}
\usepackage[utf8]{inputenc}
\usepackage{amsfonts}
\usepackage{amssymb}
\usepackage{amsmath}
\usepackage{amsthm}
\usepackage{xcolor}
\usepackage{ulem}
\usepackage{graphicx,rotating}
\usepackage{preview}
\usepackage{dashrule}

\def\b{\begin{equation}}
\def\e{\end{equation}}

\definecolor{rouge}{rgb}{.6,0,.18}
\definecolor{blueM}{rgb}{.2,0.05,0.7}

\def\ba{\begin{eqnarray}}
\def\ea{\end{eqnarray}}
\usepackage{latexsym}

\def\sech{{\rm sech}}
\def\arctanh{{\rm arctanh}}

\title{\bf  Generalized and new solutions of the NRT nonlinear Schr\"odinger equation}

\author{\\   {\bf  P. R. Gordoa, A. Pickering, D. Puertas-Centeno, E. V. Toranzo} \\ \\
             \'Area de Matem\'atica Aplicada, ESCET \\
             Universidad Rey Juan Carlos \\
             C/~Tulip\'an s/n, 28933 M\'ostoles \\
             Madrid, Spain}

\date{11 July 2024}

\begin{document}

\large

\maketitle

\thispagestyle{empty}

\begin{tabbing}

\noindent \smallskip
{\em Short title:} \= A nonlinear Schr\"odinger equation \\


\noindent \bigskip
{\em MSC2000 classification scheme numbers:} \> {}\hskip 54mm
35C06, 35C07, 35Q41, 35Q55


\end{tabbing}

\begin{abstract}

In this paper we present new solutions of the non-linear Schrödinger equation proposed by Nobre, Rego-Monteiro and Tsallis for the free particle, obtained from different Lie symmetry reductions. Analytical expressions for the wave function, the auxiliary field and the probability density are derived using a variety of approaches. Solutions involving elliptic functions, Bessel and modified Bessel functions, as well as the inverse error function are found, amongst others. On the other hand, a closed-form expression for the general solution of the traveling wave ansatz (see Bountis and Nobre) is obtained for any real value of the nonlinearity index. This is achieved through the use of  the so-called \textit{generalized trigonometric functions} as defined by Lindqvist and Drábek, the utility of which in analyzing the equation under study is highlighted throughout the paper.

\end{abstract}

\vfill

{\bf Corresponding author:} 

P. R. Gordoa. email: pilar.gordoa@urjc.es

\newpage

\setcounter{section}{0} \setcounter{equation}{0}

\section{Introduction}
In 2011 Nobre, Rego-Monteiro and Tsallis proposed a non-linear generalization of the Schrödinger equation (NRT-NLSE) inspired by nonextensive statistical mechanics \cite{NRT11,NRT12}. The NRT-NLSE has been studied for a particle subject to a uniform acceleration \cite{PLA13}, as well as under the action of infinite \cite{PLA14}, Morse type \cite{ZAN15} and Lorentzian \cite{REG20b,REG20a} potentials. Moreover, the quasi-stationary states of an arbitrary potential have been analysed \cite{TPDP13}.  This paper is devoted to the study of the free particle case, in which case the NRT-NLSE is written \cite{NRT11,BN16}
\begin{align}
\label{eq:Psi_eq}
i\hbar \frac{\partial}{\partial t}\left[\Psi(\vec{x},t) \right]	&= -\frac{1}{2-q}\frac{\hbar^2}{2m} \nabla^{2}\left[ \left(\Psi(\vec{x},t)\right) ^{2-q} \right]\, ,
\end{align}
where $\Psi(\vec{x},t)$ denotes the wave function corresponding to the system, and $q$ is the nonextensive index.
As is well-known, in the quantum formalism the probability density function (PDF, hereinafter), $\rho(\vec{x},t)$, corresponds to  the square  of the wave function's norm. However, as shown in \cite{NRT12}, the wave field $\Psi(\vec x,t)$ has to be complemented by the auxiliary field, $\Phi(\vec{x},t)$, which reduces to $\Psi^*(\vec{x},t)$ when $q \to 1$, and satisfies
\begin{align}
i\hbar \frac{\partial}{\partial t}\left[\Phi(\vec{x},t) \right]	&= \frac{\hbar^2}{2m}\left( \Psi(\vec{x},t) \right)^{1-q} \nabla^{2}\left[ \Phi(\vec{x},t)\right]\, .
\label{eq:Phi_eq}
\end{align}
Then the PDF, $\rho(\vec{x},t)$, is given by

\begin{equation}\label{eq:pdf_free_part}
\rho(\vec{x},t) = \frac{1}{2N(\Omega)} [\Psi(\vec{x},t)\Phi(\vec{x},t) + \Psi^{*}(\vec{x},t) \Phi^{*}(\vec{x},t)  ],
\end{equation}

where $N(\Omega)$ is the normalization factor, which ensures
\begin{equation}\label{eq:pdf_norm}
\int_{\Omega} \rho(\vec{x},t)\, d\vec{x} = 1 \, , \qquad ( \forall t )
\end{equation}
over an arbitrary finite volume, $\Omega$.

It is worth remarking that Eq. \eqref{eq:Psi_eq} admits solutions of the form
\begin{equation}\label{eq:q_exp_def}
\Psi(\vec x,t) = \Psi_0\, \text{exp}_q\left[i(\vec k\cdot\vec x-\omega t)\right], \quad q\in\mathbb{R},
\end{equation}
involving  the $q$-exponential function with complex argument, $\text{exp}_q(iu),$ defined as the principal value of  
\begin{equation}
\text{exp}_q(iu)=  [1+(1-q)iu]^{\frac{1}{1-q}}. 
\end{equation}
The oscillatory behavior of these solutions \eqref{eq:q_exp_def} has been highlighted via composition of (standard) trigonometric functions \cite{REG13}.

Later, a deeper study of the solutions in the traveling wave case was undertaken by Bountis et al. \cite{BN16}, restricted to one spatial dimension, in which case the wave and complementary fields reduce to $\Psi(x,t)$ and $\Phi(x,t)$, respectively. With respect to the wave field, $\Psi(x,t)$, the authors of \cite{BN16} expressed it as $\Psi( x,t)=f(v)$ with $ v=i(kx-\omega t))$, and then obtained that $f(v)$ satisfies the equation
\begin{equation}\label{eq:bountis}
\frac{df}{dv}=af^q+bf^{q-1},
\end{equation}
recovering the $q$-plane wave solutions in the particular case $b=0$. The solutions of Eq. \eqref{eq:bountis} are given in terms of a Taylor series \cite{BN16} for  $b\neq 0$; however, as far as we know, no closed form is known. Here we overcome this deficiency. In fact, we obtain a closed expression not only for the solutions of Eq. \eqref{eq:bountis} for any $q$, but also for the associated auxiliary field, $\Phi(x,t)$, and consequently for the PDF defined in Eq.\eqref{eq:pdf_free_part}. In order to do this we use the so-called \textit{Generalized Trigonometric Functions} (GTFs) initially introduced by Lindqvist \cite{Lindqvist95}, and later generalized by Drabek \cite{Drabek99} to the biparametric case. It is important to emphasize that these functions should not be confused with the generalization of trigonometric functions discussed in \cite{BOR98}, developed within the nonextensive framework and used to show the oscillatory behaviour of the complex $q$-exponential and, hence, the $q$-plane waves in Eq.\eqref{eq:q_exp_def}. Indeed, the GTFs are themselves involved in the solutions of Eq. \eqref{eq:Psi_eq} for a particle subject to an infinite potential well \cite{PLA14}. Let us briefly remark here that the $(p,q)$-sine function is defined as the inverse of  
\begin{equation} 
\text{arcsin}_{p,q} (z)=\int_0^z(1-t^q)^{-\frac1p}dt,
\end{equation} and the $(p,q)$-cosine function as the derivative of the $(p,q)$-sine. As in the standard case, the $(p,q)$-tangent is defined as their quotient.

Although the GTFs are not so well-known, some application have been found in different areas of physics, from vibration in solids \cite{Cveticanin20,Cveticanin20b,Vujkov22} to minimal length quantum theories \cite{Shababi16}. In addition, it is worth mentioning the large amount of mathematical properties which have been derived over the last two decades, from multiple-angle formulae to a variety of inequalities. Originally the study of these functions was within the context of solutions of a $p$-Laplacian eigenvalue problem \cite{Drabek99}. In this paper, we give a short introduction in Appendix \ref{sec:GTFs}, where we restrict ourselves to the information strictly necessary for an understanding of the results given here. The interested reader can find further details in, for example, \cite{Edmunds12,Bhayo12,Yin17}).

On the other hand, far beyond the travelling wave case, different approaches such as perturbation theory \cite{ZAM16}, or $q$-deformed Hilbert spaces \cite{DAC19}, have also been explored (see also the overview \cite{NOB17}). 

In the current paper, we explore the use of Lie point symmetries to obtain new solutions of the NRT-NLSE. Lie point symmetries constitute a well-known and powerful method of obtaining solutions of partial differential equations (PDEs) via reductions to ordinary differential equations (ODEs). Indeed, they have been applied to various nonlinear versions of the Schrödinger equation, giving rise to particular solutions of both theoretical and practical interest, such as soliton solutions. To the best of our knowledge, except for the case of travelling wave solutions, Lie point symmetries have not previously been used to determine solutions of the NRT-NLSE. It is our goal here to start to fill this gap, at least partially. Thus, in addition to generalizing the travelling wave results, we also obtain further new symmetry reductions of the NRT-NLSE. With respect to the physical meaning of the solutions obtained we can highlight, in some cases, a clear oscillatory behavior as evidenced by GTFs. Furthermore, the nonlinear character of the PDF of the the system is essentially captured in the generalized parameters associated to GTFs. Moreover, a solution expressed as a transformation of the Gaussian is found. A deeper study of the physical implications of our results is postponed to future publications.

The structure of the paper is as follows. In Section 2 we apply the Lie symmetry method to the NRT-NLSE system associated to a free particle in one spatial dimension. We obtain four different similarity reductions and corresponding systems of ODEs. In subsequent sections we analyze each of these systems of ODEs, thus obtaining new and generalized solutions of the NRT-NLSE and corresponding PDFs. The solutions obtained are expressed using GTFs, as well as a varity of other special functions. For the definitions and properties of special functions we refer, for example, to \cite{NIST,GR00}. In the final Section we summarize the results obtained, as well as also discussing possible future directions.

\setcounter{equation}{0}

\section{Generalized and new similarity reductions of the NRT-NLSE}

As mentioned in the Introduction, we focus in this paper on the analysis of exact solutions of the NRT-NLSE corresponding to the free particle presented in \cite{NRT11}, in one spatial dimension,
\ba
& & i\hbar \Psi_t=-\frac{1}{2-q}\frac{\hbar^2}{2m}(\Psi^{2-q})_{xx},\\
& & i\hbar \Phi_t=\frac{\hbar^2}{2m}\Psi^{1-q}\Phi_{xx}\, ,
\ea
where $\Psi\equiv \Psi({x},t)$ is the wave field, and $\Phi\equiv \Phi({x},t)$, is the auxiliary field introduced in \cite{NRT12}.   
Furthermore, we assume arbitrary $q\neq1,2$ (since for $q=1$ the problem reduces to the classical Schrödinger equation, and the problem is well-defined and can be solved for $q=2$ \cite{PLA14}). For reasons of convenience we set $b=\frac{\hbar}{2mi}$ and thus rewrite the system as
\ba
& &  \Psi_t=-\frac{b}{2-q}(\Psi^{2-q})_{xx},\label{nlsa}\\
& &  \Phi_t=b\Psi^{1-q}\Phi_{xx}.\label{nlsb}
\ea

\noindent 
We now consider the derivation of similarity reductions associated to classical Lie point symmetries \cite{BK89,O93,S89} for the system of equations (\ref{nlsa}) and (\ref{nlsb}). In order to do so we require the invariance of this system under the one-parameter Lie group of infinitesimal transformations in $(x,t,\Psi,\Phi)$ given by
\ba
& & x\rightarrow x+\epsilon\, \xi(x,t,\Psi,\Phi)+O(\epsilon^2), \\
& & t\rightarrow t+\epsilon\, \tau(x,t,\Psi,\Phi)+O(\epsilon^2), \\
& & \Psi\rightarrow \Psi+\epsilon\, \phi_1(x,t,\Psi,\Phi)+O(\epsilon^2),\\
& & \Phi \rightarrow \Phi+\epsilon\, \phi_2(x,t,\Psi,\Phi)+O(\epsilon^2),
\ea
where $\epsilon$ is the group parameter. The symmetry generator associated to
the above group of point transformations can be written as
\b
{\bf v}=\xi(x,t,\Psi,\Phi) \,\frac{\partial}{\partial x}+\tau(x,t,\Psi,\Phi)\, 
\frac{\partial}{\partial t}+
       \phi_1(x,t,\Psi,\Phi)\, \frac{\partial}{\partial \Psi}+\phi_2(x,t,\Psi,\Phi)\, \frac{\partial}{\partial \Phi}.
\e
The invariance condition leads to an overdetermined system of linear differential equations (the determining equations) for the infinitesimals $\xi, \tau$, $\phi_1$ and $\phi_2$. Once the infinitesimals have been obtained,  the similarity variables are found by solving the associated characteristic equations
\b
\frac{dx}{\xi(x,t,\Psi,\Phi)}=\frac{dt}{\tau(x,t,\Psi,\Phi)}=\frac{d\Psi}{\phi_1(x,t,\Psi,\Phi)}=\frac{d\Phi}{\phi_2(x,t,\Psi,\Phi)}.
\label{char}
\e
The infinitesimals $\xi, \tau$, $\phi_1$ and $\phi_2$ associated to the classical Lie symmetries of the system of equations (\ref{nlsa}) and (\ref{nlsb}) are found to be
\ba
\xi & =& c_3x+c_4,\label{inf1}\\
\tau & = & c_1t+c_2,\label{inf2}\\
\phi_1 & = & \frac{c_1-2c_3}{q-1}\Psi,\label{inf3}\\
\phi_2 & = & c_6\Phi+c_5x+c_7 \label{inf4}
\ea
where  $c_i, i=1\dots7$ are arbitrary constants. From the above infinitesimals we obtain four different similarity reductions. In all four of these reductions we can set $c_5=c_7=0$, since their only contribution is to add terms $\alpha+\beta x$, where $\alpha$ and $\beta$ are constant, to the expressions given below for $\Phi$ and so they make no contribution to the corresponding reduced equations. (It is clear from equation (\ref{nlsb}) that such terms can always be added to any given solution $\Phi$). Of the four reductions given here, the first provides a generalization of the previously-considered travelling wave reduction, and the other three appear to be new. These four reductions are as follow, where in each case below primes denote derivatives with respect to the reduced independent variable $z$.

\vskip 1mm

\noindent{\bf Case 1: travelling wave reduction}

The travelling wave reduction corresponds to the choice of parameters $c_1=c_3=0$, in which case we can set $c_2=1$ without loss of generality. The associated similary reduction is
\b
\Psi(x,t)=P(z),\quad \Phi(x,t)=Q(z)e^{c_6t},\quad z(x,t)=x-c_4t.
\e
The corresponding system of ODEs is
\ba
&& \frac{1}{2-q}(P^{2-q})''-\alpha P'=0,\label{stwra}\\
&& bP^{1-q}Q''+\alpha bQ'-c_6Q=0,\label{stwrb}
\ea
where $\alpha=c_4/b$. We note that the particular case of this reduction in which $c_6=0$ was considered in \cite{BN16}: the more general equation (\ref{stwrb}) seems to be new.

\vskip 1mm

\noindent{\bf Case 2: scaling reduction} 

The scaling reduction corresponds to the choice of parameters $c_3=c_1\varepsilon$ where we can set $c_2=c_4=0$ without loss of generality. The similarity variables in this case are
\b
\Psi(x,t)=P(z)t^{\frac{1-2\varepsilon}{q-1}},\quad \Phi(x,t)=Q(z)t^{-\sigma},\quad z(x,t)=xt^{-\varepsilon}
\e
where $\sigma=-c_6\varepsilon/c_3$. The corresponding system of ODEs is
\ba
&& \varepsilon z P'+\frac{2\varepsilon -1}{q-1} P=\frac{b}{2-q}(P^{2-q})'',\label{sscra}\\
&& \varepsilon z Q'+\sigma Q=-bP^{1-q} Q''.\label{sscrb}
\ea

\vskip 1mm

\noindent{\bf Case 3: \boldmath $\log(t)$ reduction} 

For the choice $c_2=c_3=0$, where we may set $c_1=1$ without loss of generality, we obtain the reduction
\b
\Psi(x,t)=P(z)t^{\frac{1}{q-1}},\quad \Phi(x,t)=Q(z)t^{-\sigma},\quad z(x,t)=x-c_4\log(t),
\e
which yields the system of ODEs
\ba
&& P-c_4(q-1)P'+\frac{b(q-1)}{2-q}(P^{2-q})''=0,\label{sr3a}\\
&& c_4Q'+\sigma Q+bP^{1-q}Q''=0.\label{sr3b}
\ea

\vskip 1mm

\noindent{\bf Case 4: \boldmath $\log(x)$ reduction}

For the choice $c_1=c_4=0$, where we may set $c_3=1$ without loss of generality, we obtain the reduction
\b
\Psi(x,t)=P(z)x^{-\frac{2}{q-1}},\quad \Phi(x,t)=Q(z)x^{-\sigma},\quad z(x,t)=t-c_2\log(x),
\e
which yields the system of ODEs
\ba 
&& P'=-b\left[c_2\left(c_2P^{1-q}P'+\frac{2}{q-1}P^{2-q}\right)'+\frac{3-q}{q-1}\left(c_2P^{1-q}P'+\frac{2}{q-1}P^{2-q}\right)\right],\label{sr4a}\\
&& Q'=bP^{1-q}\left[(\sigma+1)(c_2Q'+\sigma Q)+c_2(c_2Q'+\sigma Q)'\right].\label{sr4b}
\ea

\setcounter{equation}{0}

\section{Travelling wave reduction}

The system of ODEs obtained in the travelling wave reduction is (\ref{stwra}), (\ref{stwrb}), i.e.,
\ba
&& \frac{1}{2-q}(P^{2-q})''-\alpha P'=0,\label{twra}\\
&& bP^{1-q}Q''+\alpha bQ'-c_6Q=0,\label{twrb}
\ea
where primes denote derivatives with respect to the reduced independent variable $z$. We recall that the particular case $c_6=0$ of this reduction was studied in \cite{BN16}. We believe the more general equation (\ref{twrb}) to be new. 

Equation (\ref{twra}) can be integrated once to give
\b
P'=\alpha P^q+\beta P^{q-1} \label{twp}
\e
with $\beta$ being the arbitrary constant of integration. We note here that the particular case of equation (\ref{twp}) when $\beta=0$ was considered in \cite{NRT11} and \cite{NRT12} and that the general case when $\beta\neq0$ has been studied in \cite{BN16}. For integer $q\geq3$, we can integrate equation (\ref{twp}) to get
\b
\left(-\frac{\alpha}{\beta}\right)^{q-2}\left[\log\left(1+\frac{\beta}{\alpha}\frac{1}{P}\right)+\sum_{j=1}^{q-2} \frac{1}{jP^j}\left(-\frac{\alpha}{\beta}\right)^{-j}\right]=-\beta(z+z_0)
\e
where $z_0$ is the constant of integration, which then gives $P$ implicitly.

We now consider two different approaches to solving equation (\ref{twp}): first by mapping to ODEs of polynomial type, and secondly using generalized trigonometric functions.

\subsection{Mapping to polynomial equations}

We seek solutions of equation (\ref{twp}) via the change of variable 
\b
P(z)=u^{\frac{1}{q-1}}, 
\e
which yields
\b
u'=\alpha (q-1) u^2+\beta (q-1)u^{\frac{2q-3}{q-1}},\label{twp1}
\e
and then considering the special cases where the exponent $(2q-3)/(q-1)=n$,
a non-negative integer. The choices $n=1$ (corresponding to $q=2$) and 
$n=2$ are not permitted.

For $n=0$, corresponding to the case $q=3/2$, we get the Ricatti equation
\b
u'=\frac{1}{2}(\alpha u^2+\beta).\label{ric}
\e
This has general solution
\b
u(z)=k_0 \tanh (k_1 z+z_0)
\e
in the case in which $\alpha\beta<0$, where $z_0$ is a constant of integration and $k_0$ and $k_1$ are related to $\alpha$ and $\beta$ through $2k_1+\alpha k_0=0$ and $\beta=2k_0k_1$, which then gives
\b
P(z)=k_0^2 \tanh^2(k_1 z+z_0), 
\e
and, in the case $c_6=0$, leads to
\b
Q(z)=\frac{1}{2}Kz+\frac{K}{4k_1}\sinh[2(k_1 z+z_0)]+\tilde{K},
\e
where $K$ and $\tilde{K}$ are two arbitrary constants of integration.

When $\alpha\beta>0$ the general solution of equation (\ref{ric}) is
\b
u(z)=k_0 \tan (k_1 z+z_0),
\e
where $z_0$ is a constant of integration and where now $2k_1-\alpha k_0=0$ and $\beta=2k_0k_1$, which gives
\b
P(z)=k_0^2 \tan^2(k_1 z+z_0),
\e
and, again in the case $c_6=0$, then leads to
\b
Q(z)=\frac{1}{2}Kz+\frac{K}{4k_1}\sin[2(k_1 z+z_0)]+\tilde{K},
\e
where as before $K$ and $\tilde{K}$ are two arbitrary constants of integration. \\

For $n=3$, i.e., $q=0$, we have the equation
\b
u'=-u^2(\alpha +\beta u)
\e
whose solution is given implicitly as
\b
\frac{1}{\alpha u}-\frac{\beta}{\alpha^2}\log\left(\beta+\frac{\alpha}{u}\right)=z+z_0,
\e
where $z_0$ is the arbitrary constant of integration, or, in terms of $P(z)$,
\b
\alpha P-\beta\log\left(\beta+\alpha P\right) = \alpha^2 (z+z_0).
\e
We note that, since $P(z)$ is only determined implicitly, solving (\ref{twrb}) for $Q(z)$ is then rendered much less feasible. The same remark holds for the subsequent cases $n=4,5,6,\ldots$, for which $P(z)$ is also only determined implicitly.

We now consider the case $n=4$, which corresponds to $q=1/2$, and the equation
\b
u'=-\frac{1}{2}u^2 (\alpha+\beta u^2).
\e
The general solution of this equation is given implicitly as
\b
\frac{1}{u}+\frac{\sqrt{\alpha\beta}}{\alpha}\arctan\left(\frac{\beta u}{\sqrt{\alpha\beta}}\right)=\frac{1}{2}\alpha (z+z_0)
\e
in the case $\alpha\beta>0$, and as
\b
\frac{1}{u}+\frac{\sqrt{-\alpha\beta}}{\alpha}\arctanh\left(\frac{\beta u}{\sqrt{-\alpha\beta}}\right)=\frac{1}{2}\alpha (z+z_0)
\e
in the case $\alpha\beta<0$, where in each case $z_0$ is an arbitrary constant, and
$P(z)$ is then determined as $P=u^{-2}$.

For the cases $q=(n-3)/(n-2)$, integer $n\geq5$, equation (\ref{twp1}), i.e.,
\b
u'=(q-1)u^2(\alpha+\beta u^{n-2}),
\e
may also be solved implicitly, with $P(z)$ then being obtained as $P=u^{\frac{1}{q-1}}$.

\subsection{Solutions in terms of generalized trigonometric functions}

As an alternative approach, we consider seeking solutions of equation \eqref{twp} in
terms of generalized tangent functions. In order to do so, we seek a transformation of the form
\b
P(z)=\frac{\beta}{\alpha}(\omega(\tau))^r,\qquad z=\gamma \tau,
\e
where $r$ and $\gamma$ are constants to be determined, to an equation of the form 
\b
\frac{d\omega}{d\tau}=1+\omega^s
\e
for some exponent $s$. Balancing exponents in the resulting transformed equation, we find that there are two possible choices of $r$: $r=1/(2-q)$ and $r=1/(1-q)$.

We consider first the choice $r=1/(2-q)$. Making the change of variables
\b
P(z)=\frac{\beta}{\alpha}(w(\tau))^\frac{1}{2-q},\qquad z=\gamma \tau, \qquad
\gamma=\frac{1}{2-q}\frac{\beta^{1-q}}{\alpha^{2-q}}
\e
(where we have set $\omega=w$), we obtain the transformed equation
\b
\frac{dw}{d\tau}=1+w^{\frac1{2-q}},	
\e	
whose solutions can be expressed using generalized tangent functions (see Appendix \eqref{sec:GTFs}) as
\b
w(\tau)=\tan_{\frac1{2-q}}(\tau+\tau_0),
\e
where $\tau_0$ is an arbitrary constant of integration. We thus obtain
\b
P(z)=\frac \beta\alpha\left[ \tan_{\frac1{2-q}}\left(\frac{\alpha^{2-q}(2-q)}{\beta^{1-q}}\,(z+z_0)\right)\right]^\frac1{2-q},
\e
where we now denote the constant of integration by $z_0$ (and which is defined in terms of $\tau_0$).
The \textit{habitual} extension of Takeuchi's generalized tangent function $\tan_a$ is defined for $a>1$ which in our case imposes $\frac1{2-q}>1$, i.e., $q\in(1,2)$. However, relaxing this condition to $a>0$ (see, for example, \cite{BAR14,KAR15}) requires only that $q<2.$

We now consider the second choice $r=1/(1-q)$. Making the change of variables
\b
P(z)=\frac{\beta}{\alpha}(W(\tau))^\frac{1}{1-q},\qquad z=\gamma \tau, \qquad
\gamma=\frac{1}{1-q}\frac{\alpha^{q-2}}{\beta^{q-1}}
\e
(where we have set $\omega=W$), we obtain the transformed equation
\b
\frac{dW}{d\tau}=1+W^{\frac1{q-1}},	
\e	
whose solutions can be expressed as
\b
W(\tau)=\tan_{\frac1{q-1}}(\tau+\tau_0),
\e
where $\tau_0$ is an arbitrary constant of integration. We thus obtain
\b
P(z)=\frac \beta \alpha \left[\tan_{\frac1{q-1}}\left(\frac{\beta^{q-1}(1-q)}{\alpha^{q-2}}\,(z+z_0)	\right)\right]^\frac1{1-q},
\e
where once again we denote a redefined constant of integration by $z_0$,
and which requires $q>1$ if in $\tan_a$ we require $a=\frac{1}{q-1}>0$.

In this second case $r=1/(1-q)$ we find, given the exponent of $P$ appearing in equation \eqref{twrb}, that when $c_6=0$ it is feasible to solve for $Q$. We find
\b
Q=K_1\int e^{-\alpha \int P^{q-1}(z)dz}dz+K_2,
\e
where we can integrate to obtain
\ba
 \int P^{q-1}(z)dz&=&\frac{\beta^{q-1}}{\alpha^{q-1}}\int \tan^{-1}_{\frac1{q-1}}\left(\frac{\beta^{q-1}(1-q)}{\alpha^{q-2}}\,(z+z_0)\right)\,dz
 \nonumber \\
 &=&\frac{1}{\alpha(1-q)} \log\left(\sin_{\frac1{q-1}}\left(\frac{\beta^{q-1}(1-q)}{\alpha^{q-2}}\,(z+z_0)\right)\right),
 \ea
 and thus,
taking into account the Pythagorean-like relation \eqref{eq:pyth} and integrating by parts (see \cite{KOV19,TAK19}),
\ba
Q(z)&=&K_1 \frac{\alpha^{q-2}}{\beta^{q-1}}\bigg[\sin_{\frac1{q-1}}\left(\widetilde z\right)\cos_{\frac1{q-1}}^{\frac{2-q}{q-1}}\left(\widetilde z\right)- \widetilde z\bigg]+K_2
\ea
where  $\widetilde z=\frac{\beta^{q-1}(1-q)}{\alpha^{q-2}}\,(z+z_0)=\frac{\beta^{q-1}(1-q)}{\alpha^{q-2}}\,(x-\alpha b t+z_0).$

\begin{figure}[htb]
	\centering
	\includegraphics[width=0.5\linewidth]{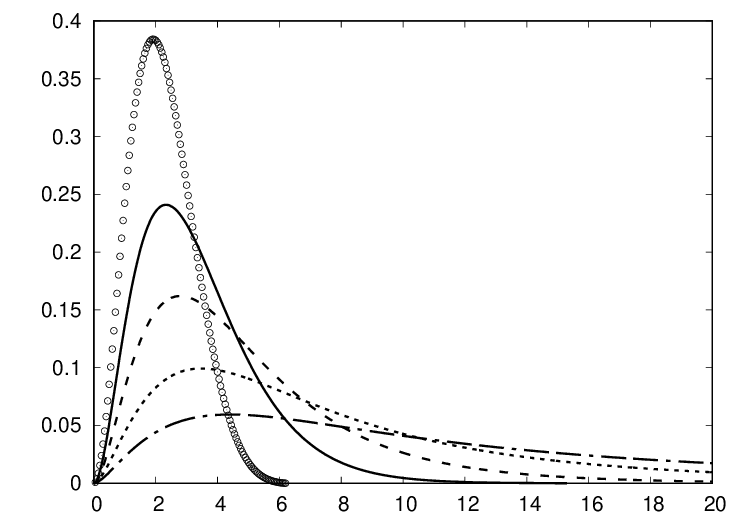}
	\caption{Graphical representation of the traveling wave probability density (Eq.~\eqref{eq:travel_density}) as a function of the variable $x$ and for fixed $t=0$ and for $q=1.9$ ($\circ$) $q=2$ (\hdashrule[0.5ex]{8mm}{0.8pt}{}), $q=2.1$ (\hdashrule[0.5ex]{8mm}{0.5pt}{3pt}), $q=2.2$ (\hdashrule[0.5ex]{8mm}{0.5pt}{1pt}) , $q=2.3$ (\hdashrule[0.5ex]{8mm}{0.5pt}{1.2mm 1pt 0.25mm 1pt}) for $\alpha=\beta=i$, $K_1=sign(q-2)$ the sign function of $q-2$ and $K_2=0$.}
	\label{fig:ondaviajera}
\end{figure}

The probability density of the system (see equation \eqref{eq:pdf_free_part}) is then
given by

\ba\nonumber
	\rho(x,t)&=&\frac {1}{N(\Omega)}\text{Re}\left\{
	\frac{\beta^{2-q}}{\alpha}\tan_{\frac1{q-1}}^\frac1{1-q} (\widetilde z)
\left( K_1\alpha^{q-2}\sin_{\frac1{q-1}} \left(\widetilde z	\right)\cos_{\frac1{q-1}}^{\frac{2-q}{q-1}}\left(\widetilde z\right)-K_1\alpha^{q-2} \widetilde{z} +
K_2\beta^{q-1}
\right)
\right\}
\\
&&\label{eq:travel_density}
	\ea
	where $z=x-\alpha b t$. For examples of the probability density in Eq.~\eqref{eq:travel_density} see Figure~\eqref{fig:ondaviajera}.

\setcounter{equation}{0}

\section{Scaling reduction} 

In the case of the scaling reduction, we obtain the system of ODEs (\ref{sscra}), (\ref{sscrb}), i.e.,
\ba
&& \varepsilon z P'+\frac{2\varepsilon -1}{q-1} P=\frac{b}{2-q}(P^{2-q})'',\label{scra}\\
&& \varepsilon z Q'+\sigma Q=-bP^{1-q} Q'',\label{scrb}
\ea
where primes denote derivatives with respect to the reduced independent variable $z$. We believe this scaling reduction of the NRT-NLSE to be new. Here we seek solutions of equation (\ref{scra}), for two particular choices of $\varepsilon$: $\varepsilon=1/(3-q)$ and $\varepsilon=1/(2(2-q))$. These correspond respectively to cases where a direct integration can be effected, or an integrating factor can be used.

\begin{figure}[htb]
	\centering
	\includegraphics[width=0.5\linewidth]{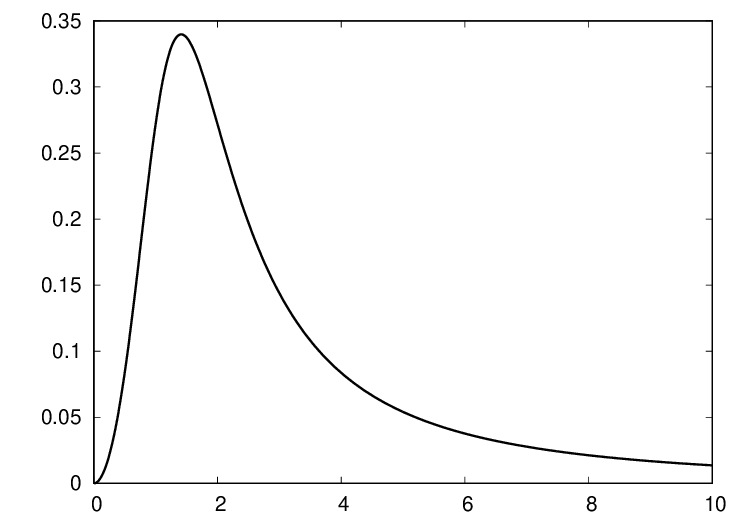}
	\caption{In this graphic we represent the probability density in Eq.~\eqref{eq:density_sec_4a} as a function of $x$ for fixed $t=1$ and for $q=2$ .}
	\label{fig:grafica_sec_4}
\end{figure}

We begin with the choice $\varepsilon=1/(3-q)$, for which equation (\ref{scra}) can be integrated once to give
\b
\frac{b}{2-q}(P^{2-q})'-\frac{1}{3-q}zP-\beta=0
\e
where $\beta$ is an arbitrary constant of integration. This equation can be alternatively written as
\b
P'=\frac{P^{q-1}}{b}\left(\beta+\frac{1}{3-q}zP\right).
\e
If $\beta=0$ this last equation becomes separable and so we can integrate again to obtain
\b
P^{1-q}=\frac{1-q}{b}\left(\frac{z^2}{2(3-q)}+\alpha\right),
\label{scap}
\e
where $\alpha$ is an arbitrary constant. We note in passing that, for $\alpha=0$, this solution will correspond to a solution obtainable through separation of variables. We also note that, for $P$ as given by (\ref{scap}), we can find the following particular solution $Q$ of equation (\ref{scrb}), for the special case $\sigma=-1$:
\b
Q(z)=z^2+2\alpha (1-q).
\e
In this case, the PDF defined in Eq.\eqref{eq:pdf_free_part} is given by
\b\label{eq:density_sec_4a}
\rho(x,t) = \frac{1}{N(\Omega)}\text{Re}\left\{\left[\frac{1-q}{b}\left(\frac{z^2}{2(3-q)}+\alpha \right)  \right]^{\frac{1}{1-q}}[z^2 + 2\alpha(1-q)]t^{\frac{q-2}{q-3}}  \right\}\, ,
\e
where $z = xt^{\frac{1}{q-3}}$. For examples of the probability density in the latter equation see Figure~\eqref{fig:grafica_sec_4}.

We now consider the choice $\varepsilon=1/(2(2-q))$. By using an integrating factor,
equation (\ref{scra}) yields the first order equation
\b
(P^{2-q})'=\frac{1}{2b}zP+\frac{P^{2-q}}{z}-\frac{\gamma}{z},
\e
where $\gamma$ is an arbitrary constant of integration. The change of variable $Y=P^{2-q}$ brings this last equation to the form
\b 
Y'=\frac{1}{2b}zY^{\frac{1}{2-q}}+\frac{Y}{z}-\frac{\gamma}{z}\label{eqY}.
\e
We now consider two subcases, correesponding to $\gamma=0$ and $\gamma\neq0$.

In the case $\gamma=0$ equation (\ref{eqY}) is a Bernoulli equation: the further change of variables $M=Y^{\frac{q-1}{q-2}}$ thus transforms it into the linear equation
\b
M'=\frac{q-1}{q-2}\frac{M}{z}+\frac{q-1}{2b(q-2)}z,
\e
whose general solution is
\b
M(z)=\frac{q-1}{2b(q-3)}z^2+K z^{\frac{q-1}{q-2}}
\e
for $q\neq3$, and 
\b
M(z)=\frac{z^2}{b}(C+\log z)
\e
for $q=3$, where in the above expressions $K$ and $C$ are arbitrary constants of integration. These solutions then respectively give $P(z)$ as defined by
\b
P^{1-q} = \frac{q-1}{2b(q-3)}z^2 + K z^{\frac{q-1}{q-2}}
\e
for $q\neq3$, and
\b
P^{-2} = \frac{z^2}{b}(\log z + C)
\e
for $q=3$. Once again, given the exponent of $P$ in equation (\ref{scrb}), we find
that, for $\sigma = 0$, we can solve for $Q$. We thus find, for $q\neq3$,
\b
Q'(z) = K_1 \left[\frac{q-1}{2b(q-3)}z^{\frac{q-3}{q-2}} + K \right]^{\frac{1}{q-1}}\, ,
\e
which then gives
\b
Q(z) = K_1 K^{\frac{1}{q-1}}z \, {}_2 F_1 \left(\frac{q-2}{q-3}, \frac{1}{1-q}; \frac{2q-5}{q-3}; \frac{q-1}{2 b K(q-3)}z^{\frac{q-3}{q-2}}   \right) + K_2\, ,
\e
$K_1$ and $K_2$ being arbitrary constants of integration. Using Eq.\eqref{eq:pdf_free_part}, the corresponding PDF is then obtained as
\ba
\rho(x,t) & = &  \frac{1}{N(\Omega)}\text{Re}\bigg\{ \left[ \frac{q-1}{2b(q-3)}z^2 + K z^{\frac{q-1}{q-2}}   \right]^{\frac{1}{1-q}}\times \nonumber \\ & &
\left[ K_1 K^{\frac{1}{q-1}}z \, {}_2 F_1 \left(\frac{q-2}{q-3}, \frac{1}{1-q}; \frac{2q-5}{q-3}; \frac{q-1}{2 b K(q-3)}z^{\frac{q-3}{q-2}}   \right) + K_2 \right]t^{\frac{1}{q-2}} \bigg\}\, ,
\ea
where $z = xt^{-\frac{1}{2(2-q)}}$. On the other hand, for $q=3$, we obtain
\b
Q'(z) = K_1(\log z +C)^{\frac{1}{2}}\, ,
\e
where $K_1$ is an arbitrarty constant, which then gives
\b
Q(z)  =  K_1\Gamma\left(\frac{3}{2}, - \log(z) - C\right) + K_2\, ,
\e
where $K_2$ is a second constant of integration and we have redefined $K_1$, and
where $\Gamma(\alpha,x)$ is the incomplete gamma function. The associated PDF, 
defined in Eq.\eqref{eq:pdf_free_part}, is
\b
\hspace{-1.75cm} \rho(x,t) = \frac{1}{N(\Omega)}\text{Re}\left\{ \left[ \frac{z^2}{b}(\log z + C)  \right]^{-\frac{1}{2}}\left[ K_1 \Gamma\left(\frac{3}{2}, -\log z - C \right) + K_2 \right]t \right\}\, ,
\e
where $z = xt^{\frac{1}{2}}$.\\

We now turn to the case $\gamma\neq 0$ of equation (\ref{eqY}). For this equation, when $q=\frac{5}{2}$, we are able to give a solution in terms of the generalized tanh function,
\begin{equation}
	Y(z) = A\, z\left[\tanh_{\frac{2}{3}}\left(\frac{B}{z}\right)\right]^{\frac{1}{3}}\, ,
\end{equation}
wherein
$A=\frac1{\sqrt{2\gamma b}}$ and $B=-3\sqrt{2b}\gamma^\frac 32$. This then gives 
\begin{equation}
P(z)=\frac {2\gamma b}{z^2}\left[\tanh_{\frac{2}{3}}\left(\frac{-3\sqrt{2b\gamma^3}}{z}\right)\right]^{\frac{-2}{3}}.
\end{equation}
Again, due to the exponent of $P$ in equation (\ref{scrb}), we are able, in the case $\sigma=0$, to find an expression for the auxiliary field $Q(z)$. On this occasion this takes the form of the quadrature
\b
Q(z)=K_1\int\sinh_{\frac23}^\frac {2}3\left( \frac{-3\sqrt{2b\gamma^3}}{z}\right)dz+K_2\, 
\e
$K_1$ and $K_2$ being two constants of integration.

\setcounter{equation}{0}

\section{The \boldmath $\log(t)$ reduction}

For this third reduction, we obtain the system of ODEs (\ref{sr3a}), (\ref{sr3b}), i.e.,
\ba
&& P-c_4(q-1)P'+\frac{b(q-1)}{2-q}(P^{2-q})''=0,\label{r3a}\\
&& c_4Q'+\sigma Q+bP^{1-q}Q''=0,\label{r3b}
\ea
where primes denote derivatives with respect to the reduced independent variable $z$.

The first of these equations can be written
\b
P''=(q-1)\frac{P'^2}{P}+\frac{c_4}{b}\frac{P'}{P^{1-q}}-\frac{1}{b(q-1)}P^q.
\e
We now make a change of variable that interchanges independent and dependent variables. We set $P'=F(P)$, which then yields the first order equation in $F(P)$,
\b
\dot{F}=(q-1)\frac{F}{P}+\frac{c_4}{b}P^{q-1}-\frac{1}{b(q-1)}\frac{P^q}{F},\label{eqf}
\e
where now $\dot{F}=dF/dP$. We consider separately the two cases where $c_4$ is zero or different from zero. 

\subsection{The case $\mathbf{c_4=0}$} 

For $c_4=0$, $z(x,t)=x$ and the solutions for $\Psi$ and $\Phi$ correspond to cases
of separation of variables. In addition, equation (\ref{eqf}) is then a Bernouilli equation. Thus, for $c_4=0$, the change of variable $G=F^2$ transforms equation (\ref{eqf}) into the linear equation
\b
\dot{G}-2(q-1)\frac{G}{P}+\frac{2}{b(q-1)}P^q=0,
\e
whose general solution is
\b
G(P)=\frac{2}{b(q-1)(q-3)}P^{q+1}+KP^{2(q-1)}\label{gep}
\e
when $q\neq 3$, and
\b
G(P)=\left(C-\frac{1}{b}\log P\right)P^4
\e
when $q=3$, where in the above expressions $K$ and $C$ are arbitrary constants of integration.

Since $P'^2=F^2=G$, we then have that, in the case $q\neq3$, $P(z)$ satisfies the first order equation
\b
P'^2=\frac{2}{b(q-1)(q-3)}P^{q+1}+KP^{2(q-1)},\label{pp2}
\e
whereas in the case $q=3$, $P(z)$ satisfies
\begin{equation}\label{eq:DEq3}
P'^2=\left(C-\frac{1}{b}\log P\right)P^4\, .
\end{equation}

Let us deal first of all with equation (\ref{eq:DEq3}), and then return later to
(\ref{pp2}). We have that
\b
P' = \pm P^2\sqrt{C - \frac{1}{b}\log P}\,,
\e
and so
\b
\int \frac{dP}{P^2\sqrt{C - \frac{1}{b}\log P}} = \pm (z + z_0)\, ,
\e
where $z_0$ is an arbitrary constant of integration.
A primitive of the LHS of this last equation can then be written in terms of the incomplete gamma function, $\Gamma(\alpha,x)$, as \cite{GR00}
\b
-i \sqrt{b}\,e^{-bC}\Gamma\left(\frac{1}{2}, \log P - bC \right) = \pm (z + z_0)\, ,
\e 
or 
\b
\Gamma\left(\frac{1}{2}, \log P - bC \right) = \pm A (z + z_0), \quad A = \frac{i}{\sqrt{b}}e^{bC}\, .
\e
Since $\Gamma\left(\frac{1}{2}, x \right) = \sqrt{\pi}\left(1-\text{erf}\left(\sqrt{x}\right)\right)$, we have
\begin{align*}
\sqrt{\pi}\left[1-\text{erf}\left(\sqrt{\log P - bC} \right)\right] & =  \pm A (z + z_0) \\
\sqrt{\log P - bC} & = \text{erf}^{-1}\left[1 \mp \frac{A}{\sqrt{\pi}} (z + z_0)\right] \, ,
\end{align*}
which finally gives
\b\label{eq:erf_sols}
P(z) = e^{\left\{\text{erf}^{-1}\left[1 \mp \frac{A}{\sqrt{\pi}} (z + z_0)\right]\right\}^2 + bC }\, .
\e

It is worth mentioning that a family of  probability densities similar to Eq. \eqref{eq:erf_sols} appears as the minimizers of a generalization of the Stam inequality. In particular, it has been shown that these probability densities minimize Shannon entropy when a \textit{generalized Fisher information} is fixed (see Zozor et al \cite{ZOZ17}). In addition, and remarkably, equation \eqref{eq:erf_sols} corresponds precisely to a transformation of the Gaussian density very similar to the Sundman transformation discussed below, but imposing a conservation-law for the probability in any differential interval \cite{Puertas19,ZOZ17}.

The corresponding auxiliary function $Q$ is difficult to compute when $\sigma\neq0$; for $\sigma=0$, it takes a trivial linear form $Q(z)=K_1z+K_2$ (with $K_1$, $K_2$ being arbitrary constants).

We now return tor equation (\ref{pp2}). We begin by rewriting this equation as
\b
P'^2=P^{2(q-1)}+A P^{q+1}, \label{eqp2}
\e
where derivatives are now with respect to the variable $Z=\sqrt{K}z$ and $A=\frac{2}{Kb(q-1)(q-3)}$.

Let us remark first of all that the change of variables
\b
P(Z)=A^{\frac{1}{q-3}}(R(\tau))^{\frac{1}{2-q}},\qquad
Z=\frac{1}{2-q}A^{\frac{2-q}{q-3}}\tau,
\e
gives 
\begin{equation}
\left(\frac{dR}{d\tau}\right)^2=1+R^{\frac{q-3}{q-2}},
\end{equation}
whose solution can be written using biparametric generalized trigonometric functions as
\begin{equation}
R(\tau)=\sinh_{2,\frac{q-3}{q-2}}(\tau+\tau_0).
\end{equation}
We thus obtain
\begin{equation}
P(z)=A^{\frac 1{q-3}} \left[\sinh_{2,\frac{q-3}{q-2}}\left((2-q)\sqrt{K}A^{\frac {q-2}{q-3}}\,(z+z_0)\right)\right]^{\frac1{2-q}},
\end{equation}
where we denote a redefined constant of integration by $z_0$. Again it is difficult to
compute the auxiliary field when $\sigma\neq0$, and for $\sigma=0$ it has the trivial linear form $Q(z)=K_1z+K_2$ ($K_1$, $K_2$ arbitrary constants).

We may alternatively seek solutions of (\ref{eqp2}) by transformating to a broader class of ordinary differential equations, again with known general solutions. We now look for transformations that map equation (\ref{eqp2}) onto a first order nonlinear ODE of second degree of the form
\b
V'^2={\cal P}(V)\label{veq0}
\e
in new dependent and independent variables $V=V(\tau)$, where primes now denote derivatives with respect to $\tau$, and where ${\cal P}(V)$ is a polynomial in V of degree less than or equal to four. The solutions of this equation may be defined in terms of elliptic functions, a necessary condition for which is that ${\cal P}(V)$ be of degree three or four, or elementary functions, e.g., when ${\cal P}(V)$ is of degree one or two. We thus consider, as in \cite{KS14}--\cite{GP21}, the combined Sundman power-type transformation 
\b
P=V^{\varepsilon},\quad dZ=P^{\delta} \varepsilon d\tau, \label{sundam}
\e
where $\delta$ and $\varepsilon$ are real numbers and $\varepsilon\neq 0$ (we note that the Sundman transformation was originally proposed in \cite{Sud13}; see also \cite{DMS94}). Applying the transformation (\ref{sundam}) to equation (\ref{eqp2}) yields
\b
V'^2=V^{\varepsilon(q-1+2\delta)+2}(A+V^{\varepsilon (q-3)}).\label{veq}
\e
We seek choices of $\varepsilon$ and $\delta$ such that this equation has the form
\b
V'^2= A V^m+V^n,
\label{Vmn}
\e
for integers $m,n\in\{0,1,2,3,4\}$. We may assume $n>m$ since, if $n<m$, the change of variable $V=1/W$ gives $W'^2= A V^{4-m}+V^{4-n}$ where the new exponents satisfy $4-n>4-m$. (We note also that $n=m$ requires $\varepsilon (q-3)=0$, which is not permitted). We thus obtain the following equations for $\varepsilon$ and $\delta$:
\ba
\varepsilon (q-3) & = & n-m, \\
2\delta (n-m) & = & 2m(q-2)-n(q-1)-2(q-3).
\ea
We are then led to ten equations of the form (\ref{Vmn}), for: $m=0$, $n=1,2,3,4$; $m=1$, $n=2,3,4$; $m=2$, $n=3,4$; and $m=3$, $n=4$.

For example, for the choice $\varepsilon=1/(q-3)$ and $\delta=2-q$, we obtain $m=1$ and $n=2$, so equation (\ref{veq}) is
\b
V'^2=AV+V^2,
\e
whose general solution is given by
\b
V=-\frac{A}{2}(1+\cosh(\tau+\tau_0)), \quad 
\e
whereas for the choice $\varepsilon=1/(q-3)$ and $\delta=(1-q)/2$  we obtain $m=2$ and $n=3$, and equation (\ref{veq}) is
\b
V'^2=AV^2+V^3,
\e
with general solution
\b
V=-A \left[\sech \left(\frac{1}{2} \sqrt{A}(\tau+\tau_0)\right)\right]^2,  \quad 
\e
$\tau_0$ in each of the above expressions being an arbitrary constant of integration.
As a further example, but this time solved in terms of elliptic functions, we note that for the choice $\varepsilon=3/(q-3)$ and $\delta=(9-5q)/6$  we obtain $m=0$ and $n=3$, and equation (\ref{veq}) is
\b
V'^2=A+V^3,
\e
which has general solution given in terms of the Weierstrass $\wp$-function,
\b 
V=4\wp(\tau+\tau_0;0,-A/16), 
\e
with $\tau_0$ again being an arbitrary constant of integration. In this way we obtain solutions of equations (\ref{Vmn}), related by the combined Sundman power-type transformation (\ref{sundam}) to (\ref{eqp2}), and so to (\ref{pp2}).

\subsection{The case $\mathbf{c_4\neq 0}$} 

For $c_4\neq 0$ we can make in (\ref{eqf}) the change of variables
\b
F(P)=-c_4P^{q-1}w(y),\quad y=-P/b,
\e
to obtain an equation in $w(y)$ (here we use $w_y$ to denote the derivative),
\b
ww_y=w+Ay^{2-q}, \label{eqw}
\e
where $A=\frac{(-b)^{2-q}}{c_4^2(q-1)}$. Equation (\ref{eqw}) is an Abel equation. We can find the solutions of this equation in parametric form for three particular choices of $q$: $q=3$, $q=4$ and $q=5/2$. In the case $q=3$ we get the solution
\b
y=Kf^{-1}e^{\mp \tau^2},\quad w=Kf^{-1}\left[e^{\mp \tau^2}\pm 2\tau f\right],\quad f=\int e^{\mp \tau^2} d\tau -C,
\e
where $A=\mp 2K^2$. For $q=4$  and $q=5/2$ the general solution can be expressed parametrically in terms of Bessel functions \cite{PZ03}. For $q=5/2$ the solution is
\b
y=a\tau^{-4/3}Z^{-2}U_1^2,\quad w=a\tau^{-4/3}Z^{-2}U_2
\e
where $A=\pm \frac{1}{3}a^{3/2}$, whereas for $q=4$ we obtain
\b
y=2a\tau^{4/3}Z^2U_2^{-1}, \quad w=\pm 3a\tau^{-2/3} Z^{-1}U_2^{-1}U_3
\e
where $A=-36a^3$. In the above two solutions $Z=C_1 J_{1/3}(\tau)+C_2Y_{1/3}(\tau)$ for the upper sign and $Z=C_1 I_{1/3}(\tau)+C_2K_{1/3}(\tau)$ for the lower sign (where $
J_{1/3}(\tau)$ and $Y_{1/3}(\tau)$ are Bessel functions and 
$I_{1/3}(\tau)$ and $K_{1/3}(\tau)$ are modified Bessel functions), and $U_1=\tau Z'+\frac{1}{3}Z$, $U_2=U_1^2\pm\tau^2Z^2$ and $U_3=\pm \frac{2}{3}\tau^2Z^3-2U_1U_2$.

\setcounter{equation}{0}

\section{The \boldmath $\log(x)$ reduction}

For the last of our reductions, we obtain the system of ODEs (\ref{sr4a}), (\ref{sr4b}), i.e.,
\ba 
&& P'=-b\left[c_2\left(c_2P^{1-q}P'+\frac{2}{q-1}P^{2-q}\right)'+\frac{3-q}{q-1}\left(c_2P^{1-q}P'+\frac{2}{q-1}P^{2-q}\right)\right],\label{r4a}\\
&& Q'=bP^{1-q}\left[(\sigma+1)(c_2Q'+\sigma Q)+c_2(c_2Q'+\sigma Q)'\right].\label{r4b}
\ea

Let us first of all remark that in the case $c_2=0$, $z(x,t)=t$, and the system (\ref{r4a}), (\ref{r4b}) reduces to
\ba
&&P'=2b\frac{q-3}{(q-1)^2}P^{2-q},\\
&&Q'=b\sigma(\sigma+1)P^{1-q}Q,
\ea
whose general solution is given by
\ba
&&P(z)=\left(2b\frac{q-3}{q-1}(z+z_0)\right)^{\frac{1}{q-1}},\\
&&Q(z)=K\left(z+z_0\right)^{\frac{\sigma(\sigma+1)(q-1)}{2(q-3)}},
\ea
where $z_0$ and $K$ are arbitrary constants of integration. The solutions for $\Psi$ and $\Phi$ then correspond to cases of separation of variables, see, e.g., \cite{BN16}.
The associated PDF is
\b
\rho(x,t) = \frac{1}{N(\Omega)}\text{Re}\left[K\left(2b\frac{q-3}{q-1} \right)^{\frac{1}{q-1}}(t+t_0)^{\frac{1}{q-1}+\frac{\sigma(\sigma+1)(q-1)}{2(q-3)}}x^{-\frac{2}{q-1}-\sigma}  \right]\, ,
\e
where, as stated before, for this case $z = t$.

We now consider the case $c_2\neq0$. We begin by rewriting equation (\ref{r4a}) as
\b
	P P''+ (1-q)(P')^2 + \frac{1}{c_2}\left(\frac{7-3q}{q-1}\right) P P'+\frac{1}{bc_2^2}P^{q}P'+\frac{2}{c_2^2}\left(\frac{3-q}{(q-1)^2}\right)P^2  = 0
\e
In this autonomous equation we may set $w(P)=P'$, which then gives
\b
\dot w+ (1-q)\frac{w}{P} + \frac{1}{c_2}\left(\frac{7-3q}{q-1}\right)  +\frac{1}{bc_2^2}P^{q-1}+\frac{2}{c_2^2}\left(\frac{3-q}{(q-1)^2}\right)
\frac{P}{w}  = 0,
\e
where $\dot{}$ denotes a derivative with respect to $P$.
Here we consider the two particular cases $q=3$ and $q=7/3$,
which respectively yield a linear equation and an Abel equation.
For $q=3$ we get
\b
\dot w-2\frac{w}{P} - \frac{1}{c_2} +\frac{1}{bc_2^2}P^{2}=0,
\e
which has the general solution
\b
w(P) =  K P^2 -\frac{1}{bc_2^2}P^3 - \frac{1}{c_2}P,
\e
with constant of integration $K$. Solutions of \eqref{r4a} for $q=3$ are then obtained via
\b
-bc_2^2\int \frac{dP}{P^3 - K b c_2^2 P^2 + bc_2P} = z + z_0\, .
\e

In the case $q = \frac{7}{3}$, one has 
\b
w\,\dot w = \frac{4}{3P}w^2 - \frac{1}{b\, c_2^2}P^{\frac{4}{3}} w - \frac{3}{4c_2^2} P\, .
\e
The substitution $s(P) = wP^{-\frac{4}{3}}$ then gives an Abel equation of second kind,
\b\label{eq:Abel_1}
s\, \dot s = -\frac{1}{bc_2^2} s - \frac{3}{4c_2^2} P^{-\frac{5}{3}}\, .
\e 
Finally, the change of variable $\xi = - \frac{1}{bc_2^2} P$, transforms Eq. \eqref{eq:Abel_1} into the canonical form,
\b\label{eq:Abel_2}
s \frac{ds}{d\xi} - s = -\left(\frac{3}{4b^{\frac{2}{3}}(c_2^2)^{\frac{5}{3}}}\right) \xi^{-\frac{5}{3}}\, .
\e

\section{Conclusions}

In this paper a set of new and generalized solutions of the NRT Nonlinear Schrödinger equation for the free particle have been obtained using, to the best of our knowledge for the first time, the Lie symmetry method. These new and generalized solutions have been given, for both the wave and auxiliary fields, in terms of the Generalized Trigonometric Functions (GTFs) introduced by Lindqvist\cite{Lindqvist95} and Drábek \cite{Drabek99}, as well as other special functions, for example, elliptic and Bessel functions. This has allowed us to obtain, in some cases, closed-form expressions for the probability density of the system. In particular, for the travelling wave reduction, the probability density can be expressed in a simple way using GTFs for any real value of the NRT-NLSE nonlinearity parameter $q\neq1,2$. This fact highlights the expected oscillatory behavior of the solution. 

Future work will have the aim of further studying the physical interpretation of the solutions obtained, as well as of obtaining new or generalized solutions of the NRT-NLSE for a particle subject to a non-zero potential.

\setcounter{equation}{0}

\appendix

\section{A short introduction to generalized trigonometric functions}\label{sec:GTFs}
The $(p,q)$-generalized trigonometric functions, $\sin_{p,q},\cos_{p,q},\tan_{p,q}$ and their respective hyperbolic counter-parts were first defined, as far as we know, by Lindqvist \cite{Lindqvist95}, and Drábek and Manásevich \cite{Drabek99}. For example, the $(p,q)$-sine function is defined as the inverse of 
\begin{equation} \label{arcsin_def}
\text{arcsin}_{p,q} (z)=\int_0^z(1-t^q)^{-\frac1p}dt=z\,_2F_1\left(\frac 1p,\frac1q;1+\frac 1q;z^q\right),  \quad p,q>1,
\end{equation}
which as we see can also be expressed using the Gaussian hypergeometric function
(see, for example \cite{Edmunds12,Bhayo12}). The cosine function is defined as the derivative of the sine function 
\begin{equation}
\cos_{p,q}(z)=\frac{d}{dz}\sin_{p,q}(z).
\end{equation}
From these definitions, and applying the inverse function rule, the following Pytheagorean-like identity is obtained:
\begin{equation}\label{eq:pyth}
\cos_{p,q}^p(z)+\sin_{p,q}^q(z)=1. 
\end{equation}
The corresponding tangent function is defined as 
\begin{equation}
\tan_{p,q}=\frac{\sin_{p,q}(z)}{\cos_{p,q}(z)}.
\end{equation}

In the case $q=p$ the notation $\tan_{p,p}=\tan_p$ is usually adopted. It then follows that 
\begin{equation}
\frac{d}{dz}\tan_{p}(z)=1+\tan_{p}^p(z).
\end{equation}
In the hyperbolic case one has
\begin{equation} \label{arcsinh_def}
\text{arcsinh}_{p,q} (z)=\int_0^z(1+t^q)^{-\frac1p}dt=z\,_2F_1\left(\frac 1p,\frac1q;1+\frac 1q;-z^q\right), 
\end{equation}
and it follows that
\begin{equation}
\cosh^p_{p,q}(z)-\sinh^q_{p,q}(z)=1,\qquad (\tanh_p(z))'=1-\tanh_p^p(z)
\end{equation}
where
\begin{equation}
\cosh_{p,q}(z)=(\sinh_{p,q}(z))',\qquad \tanh_{p,q}(z)=\frac{\sinh_{p,q}(z)}{\cosh_{p,q}(z)}.
\end{equation}

\begin{figure}[htb!]
	\centering
	\includegraphics[width=0.5\linewidth]{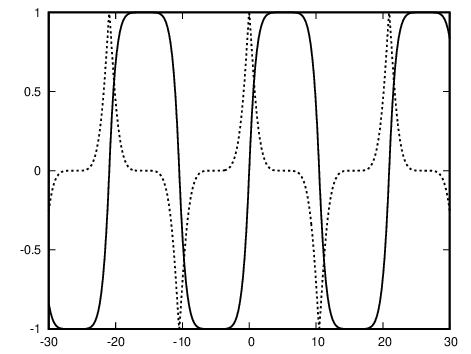}
	\caption{In this graphic we represent generalized sine (continuous) and cosine (dashed) functions when $p=q=1.2$}
	\label{fig:grafica_GTFs_1.2}
\end{figure}

\begin{figure}[htb!]
	\centering
	\includegraphics[width=0.5\linewidth]{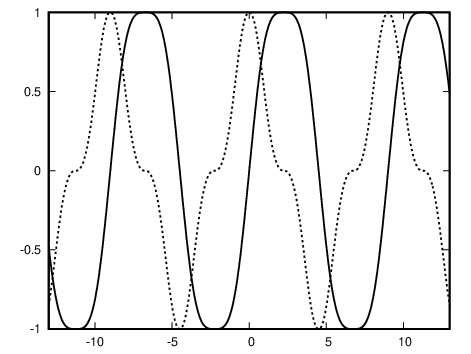}
	\caption{In this graphic we represent generalized sine (continuous) and cosine (dashed) functions when $p=1.4$ and $q=2.2$}
	\label{fig:grafica_GTFs_1.4_2.2}
\end{figure}

\begin{figure}[htb!]
	\centering
	\includegraphics[width=0.5\linewidth]{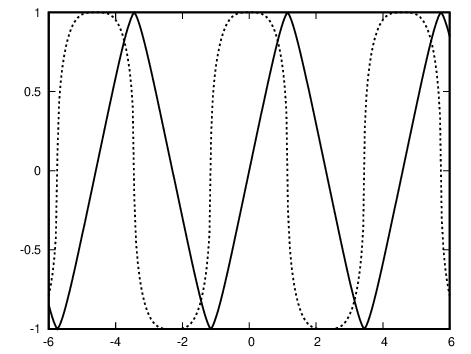}
	\caption{In this graphic we represent generalized sine (continuous) and cosine (dashed) functions when $p=q=3.5$}
	\label{fig:grafica_GTFs_3.5}
\end{figure}

Finally, we remark the recently obtained duality relations (see references \cite{MIY21,MIY22}):
	\begin{equation}\label{eq:duality}
	\sinh_{p,q}x=\frac{\sin_{r,q} x}{\cos_{r,q}^{r/q} x},\qquad \cosh_{p,q}x=\frac{1}{\cos_{r,q}^{r/p} x},
	\end{equation}
and
	\begin{equation}\label{eq:duality2}
	\sin_{p,q}x=\frac{\sinh_{r,q} x}{\cosh_{r,q}^{r/q} x},\qquad \cos_{p,q}x=\frac{1}{\cosh_{r,q}^{r/p} x},
	\end{equation}
	where in each case 
\begin{equation}
\frac1p+\frac1r=1+\frac 1q.
\end{equation}

or examples of these functions see Figures~\eqref{fig:grafica_GTFs_1.2},~\eqref{fig:grafica_GTFs_1.4_2.2} and~\eqref{fig:grafica_GTFs_3.5}.

\newpage 
\section*{Acknowledgements}

We gratefully acknowledge the following financial support:
Project PID2020-115273GB-I00 and Grant RED2022-134301-T funded by 
MCIN/ AEI/ 10.13039/501100011033.
PRG and AP also gratefully acknowledge financial support from the
Universidad Rey Juan Carlos as members of the Grupo de investigaci\'on de alto rendimiento DELFO.

\end{document}